\documentstyle[preprint,aps]{revtex}

\begin{document}

\title{Non--standard Construction of Hamiltonian Structures}

\author{Andr\'es Gomberoff$^{1}$ and Sergio A. Hojman$^{2}$ }

\address{$^1$Centro de Estudios Cient\'{\i}ficos de Santiago, Casilla
16443, Santiago, Chile \\
$^2$Departamento de F\'{\i}sica, Facultad de Ciencias,
Universidad de Chile, Casilla 653, Santiago, Chile}

\maketitle

\begin{abstract}
Examples of the construction of Hamiltonian structures for dynamical
systems in field theory (including one reputedly non--Hamiltonian problem)
without  using Lagrangians, are presented. The recently developed method
used requires the knowledge of one constant of the motion of the system
under consideration and one solution of the symmetry equation.
\end{abstract}

\section{Introduction}

Hamiltonian theories have been widely used in almost all areas of physics. The
usual approach consists in constructing the momenta (canonically conjugate  to
the coordinates) and the Hamiltonian starting from a Lagrangian formulation of
the system under consideration.  Nevertheless, there has been increasing
interest in studying non standard procedures to produce Hamiltonian structures
starting from the equations of motion only, without using a Lagrangian
\cite{morri1}--\cite{sh-mr2}. These approaches usually deal with systems which
are naturally described in terms of non--canonical variables and the
Hamiltonian formulation is consequently written using non--canonical Poisson
matrices, which in many cases are singular. Due to the lack of a general
procedure to generate Hamiltonian structures from scratch, in most instances
the results have been obtained by extremely inspired guesswork.

The purpose of this article is to present examples of the application of a
newly devised method \cite{sh,sh1} which constitutes a {\it general technique}
for the construction of Hamiltonian structures for dynamical systems. We
discuss systems described by non--linear equations in field theory and also
some linear equations such as the time dependent Schr\"odinger equations and
the heat equation, which according to the usual belief, is clearly
non--Hamiltonian \cite{salmon}.

In Section 2 we give a brief outline of the method and the rest of the
Sections are devoted to the construction of the examples.

\section{Brief Outline of the Method}

The contents of this Section summarize partially the results of \cite{sh}.
We use classical mechanical notation nevertheless, the results may be
easily applied to field theory as well, as we will see at the end of this
section.

Consider a dynamical system defined by equations which have been cast in first
order form,
\begin{equation}
\frac{dx^a}{dt} = f^a(x^b)\ \,\ \ \ \ \ \ \ \ \ \ a,b = 1,........,N\ .
\label{eqmo}
\end{equation}

A Hamiltonian structure for it consist of an antisymmetric matrix,
$J^{ab}(x^c)$ and a Hamiltonian $H(x^c)$ such that $J^{ab}$ is the Poisson
bracket for the variables $x^{a}$ and $x^{b}$ (which are non--canonical in
general) and $H$ is the Hamiltonian for system (\ref{eqmo}). In addition
to its antisymmetry, the matrix $J^{ab}$ is required to satisfy the Jacobi
identity and to reproduce, in conjunction with the Hamiltonian $H$,
the dynamical equations (\ref{eqmo}), i.e.,
\begin{equation}
{J^{ab}}_{,d} J^{dc} + {J^{bc}}_{,d} J^{da} + {J^{ca}}_{,d}  J^{db}
\equiv
0\ ,
\label{ji}
\end{equation}
and,
\begin{equation}
J^{ab}  \frac{\partial H}{\partial x^b} = f^a\ .
\label{hami}
\end{equation}

It has been proved \cite{sh} that one solution to the problem of finding
a Hamiltonian structure for a given dynamical system is provided by one constant
of the motion which may be used as the Hamiltonian $H$, and a symmetry vector
$\eta^{a}$ which allows for the construction of a Poisson matrix $J^{ab}$. The
constant of the motion and the symmetry vector satisfy,
\begin{eqnarray}
{\cal L}_f H &=& 0 \label{cc} \ ,\\
(\partial_t + {\cal L}_f) \eta^a &=& 0  \label{sym} \ ,
\end{eqnarray}
respectively, where ${\cal L}_f$ is the Lie derivative along $f$ (for a
definition,
see \cite{wein}, for instance).  
In addition, it is required that the deformation $K$ of $H$ along $\eta^{a}$,
\begin{equation}
K \equiv \frac{\partial H}{\partial x^a}\eta^a = {\cal L}_\eta H
 \ ,
\label{deform}
\end{equation}
be non-vanishing.
The Poisson matrix $J^{ab}$ is constructed as the antisymmetrized product
of the flow vector $f^{a}$ and the ``normalized'' symmetry vector
$\eta^{b}/K$,
\begin{equation}
J^{ab} =\frac{1}{K}(f^a \eta^b - f^b \eta^a)  \ .
\label{ans}
\end{equation}

The Poisson matrix so constructed has rank 2 and it is, therefore, singular
in many instances.  Adding together two Poisson matrices constructed according
to (\ref{ans}) will not increase its rank. It will just redefine the
symmetry vector used to construct it. One method to increase the rank
of such a Poisson matrix is presented in \cite{sh}.

Let us now deal with  systems with infinitely many  degrees of freedom defined
by some field $\phi(x,t)$, where $x$ denotes the coordinates of a point in
space.
The dynamical equation (\ref{eqmo}) now is,
\begin{equation}
\dot{\phi}(x,t) = F[\phi,x] \ ,
\end{equation}
where $F$ is a functional of $\phi$ for every point $x$ in space. All the above
discussion remains valid replacing $x^a$ by $\phi(x)$, tensors
$T^{ab\ldots c}(x^{a})$ by functionals $\Theta$ which depend on some spatial
coordinates, $\Theta[\phi,x,y\ldots,z]$ and partial derivatives
$\partial/\partial x^{a}$ by functional derivatives $\delta/\delta\phi(x)$.
Details are given in \cite{shepherd}.

\section{Heat Equation}

Let us consider the heat equation,
\begin{equation}
u_{t} = u_{xx}  \ .
\label{heat}
\end{equation}
It is easy to see that
\begin{equation}
\eta(x) = \Delta \  ,
\end{equation}
is a symmetry transformation for it, where $\epsilon$ is any real
number.

If we assume periodic boundary conditions in $x \in [-a/2,a/2]$ then
the quantity
\begin{equation}
H = \int_{-a/2}^{a/2} u dx
\label{heatham}
\end{equation}
is conserved.

The deformation of $H$ along $\eta$ is given by
\begin{equation}
{\cal L}_{\eta} H = \int_{-a/2}^{a/2} dx\frac{\delta H}{\delta u(x)}
\eta(x)  = \Delta \int_{-a/2}^{a/2} dx =
\Delta a  \ ,
\end{equation}
where all the integrals are taken from $-a/2$ to $a/2$.

We can now construct a Poisson matrix associated with the Hamiltonian
(\ref{heatham}) for equation (\ref{heat}),
\begin{equation}
J(x,y) = \frac{u_{xx}(x) - u_{yy}(y)}{a}   \ .
\end{equation}

Let us explicitly show that $J^{ab}$ together with $H$ provide a
Hamiltonian formulation for the heat equation. In fact,
\begin{eqnarray*}
[u(x),H] &=& \int_{-a/2}^{a/2} dy  J(x,y)\frac{\delta H}{\delta u(y)} \\
         &=& \frac{1}{a}u_{xx}(x)\int_{-a/2}^{a/2} dy  -
             \frac{1}{a}\int_{-a/2}^{a/2}u_{yy}(y) dy \\
         &=& u_{xx}(x)   \ .
\end{eqnarray*}
So we see that,
\begin{equation}
\dot{u}=[u,H]   \ ,
\end{equation}
as required.
It is worth noting that according to folk tradition, this equation cannot
be endowed with a Hamiltonian structure. In Salmon's words \cite{salmon}:
``By anyone's definition, (0.1) is non--Hamiltonian''. In his paper, (0.1)
is the heat equation subject to periodic boundary conditions.

\section{Time Dependent Schr\"odinger Equations}

Consider the following equations of motion
\begin{equation}
\psi_t =i\left(\psi_{xx}+V(x,t) \psi\right)  \equiv f \ ,
\label{sch}
\end{equation}
and its complex conjugate
\begin{equation}
{\psi^*}_t =-i\left({\psi^*}_{xx}+V(x,t)\psi^*\right)  \equiv f^*\ .
\label{sch*}
\end{equation}
Note that we discuss the case of a time dependent potential. We use one
dimensional notation for simplicity only, our results hold irrespective of
the dimensionality of space.

It is a straightforward matter to realize that multiplication of the
variables $\psi$ and $\psi^*$ by two different constants, $(1+\lambda)$ and
$(1+\mu \lambda)$ respectively, constitutes a symmetry
transformation for the Schr\"odinger equations (\ref{sch}) and
(\ref{sch*}).
The infinitesimal version of this transformation is
\begin{equation}
\eta = \psi\  ,
\label{schsymm}
\end{equation}
and
\begin{equation}
\eta^* = \mu \psi^*\  .
\label{schsymm*}
\end{equation}
The usual probability conservation statement means that
\begin{equation}
H=\int \psi^*(x) \psi(x) dx \ ,
\label{schham}
\end{equation}
is a conserved quantity as it may be easily proved. Its deformation $K$
along the symmetry vector defined by (\ref{schsymm}) and (\ref{schsymm*})
may be written in terms of variational derivatives as
\begin{equation}
K \equiv \int \frac{\delta H}{\delta \psi(x)} \eta(x) dx  +
\int \frac{\delta H}{\delta \psi^*(x)} \eta^*(x) dx  \ ,
\label{schk1}
\end{equation}
and a straightforward calculation yields
\begin{equation}
K= (1 +\mu) H  \neq 0 \ ,
\label{schk2}
\end{equation}
for any $\mu$ such that $1 + \mu \neq 0$ .
The Poisson structure is defined by
\begin{equation}
[ \psi(x) , \psi(y)] =\frac{1}{K} ( f(x) \eta(y) - \eta(x)
f(y)
 )\ ,
\label{schpb1}
\end{equation}
\begin{equation}
[\psi(x) , \psi^*(y) ] = \frac{1}{K} ( f(x) \eta^*(y) -
\eta(x)\
f^*(y))\ ,
\label{schpb2}
\end{equation}
and
\begin{equation}
[\psi^*(x),\psi^*(y)]=\frac{1}{K}(f^*(x)\eta^*(y)-
\eta^*(x) f^*(y))\ .
\label{schpb3}
\end{equation}
We have thus constructed a family of Poisson structures which depend on
the parameter $\mu$ for a given Hamiltonian $H$. Note that in spite of the
time dependent character of the Schr\"odinger equations (\ref{sch}) and
(\ref{sch*}) the Hamiltonian (\ref{schham}) is conserved, while the usual
Hamiltonian contains the time dependent potential $V(x,t)$ and it is not
conserved. On the other hand, our Poisson structures are time dependent.
This is, as far as we know, a novel feature for Poisson matrices, which
means that, even in the case of a regular matrix, the Hamiltonian
structure
is not derivable from a Lagrangian.

\section{Korteweg--De Vries Equation}

The equation of motion is
\begin{equation}
u_t=  -  u u_x -  u_{xxx}  \equiv   f \  .
\label{kdv}
\end{equation}
It is not difficult to see that symmetry transformation for it is given by
\cite{sh}
\begin{equation}
\eta   =     ( - 2 u - x u_x + 3 t ( u u_x + u_{xxx} ) ) \ .
\label{kdvsymm}
\end{equation}
In fact, to prove that (\ref{kdvsymm}) is a symmetry transformation for
the KdV
equation, it is enough to check that $\eta$ satisfies the symmetry
equation
(\ref{sym}) with $f$ defined by (\ref{kdv}).

To get a Hamiltonian structure for the KdV equation we need to construct
constants of the motion which be non trivially deformed by $\eta$. In
\cite{sh}
the energy
\begin{equation}
H_1 =  \int u^2 dx \  ,
\label{kdvh1}
\end{equation}
was used as a constant of motion to complete the Hamiltonian structure.
Note
that the deformation $K_1$ of $H_1$ along $\eta$ is non--vanishing
\begin{equation}
K_1 \equiv \int \frac{\delta H_1}{\delta u(x)} \eta(x) dx  = -  3
 H_1 \ .
\label{kdvk1}
\end{equation}
The Poisson structure ${J_1}(x,y)$ is given by
\begin{equation}
J_1 ( x , y )  = \frac{1}{K_1}  (  f(x) \eta(y) - f(y) \eta(x)
)\  .
\label{kdvj1}
\end{equation}
Consider now $H_2$
\begin{equation}
H_2  =  \int  ( - \frac{{u_x}^2}{2} + u^3) dx \ ,
\label{kdvh2}
\end{equation}
which is also conserved. Its deformation $K_2$ along $\eta$ is
\begin{equation}
K_2 \equiv \int \frac{\delta H_2}{\delta u(x)} \eta(x)dx  = -5
\
 H_2\ .
\label{kdvk2}
\end{equation}
The Poisson structure ${J_2}(x,y)$ is given by
\begin{equation}
J_2 ( x , y )  = \frac{1}{K_2} (   f(x) \eta(y) - f(y) \eta(x) \
)\  .
\label{kdvj2}
\end{equation}
We have been able to construct two different Hamiltonian structures based
on
one symmetry vector $\eta$ and two conserved quantities $H_1$ and $H_2$ as
the Hamiltonians of each of the structures. The Poisson structures $J_1$
and $J_2$ are built as the antisymmetric product of the evolution vector
$f$
and the symmetry vector $\eta$ normalized using the deformations $K_1$ and
$K_2$ of $H_1$ and $H_2$, respectively. Note that a different but closely
related scheme exists in which no normalization of the antisymmetric
product
is needed provided that the Hamiltonians $ - (1/3) \log H_1$ and
 $ - (1/5) \log H_2$ be used instead of $H_1$ and $H_2$ respectively. In
 this
case the Poisson matrix for both Hamiltonian structures is exactly the
same one.

A similar construction may be performed with the rest of the constants of the 
motion which belong to this family, as they appear, for instance, in \cite{lax}. 
We have then constructed a set of infinitely many Hamiltonian structures for the 
KdV based on one symmetry transformation and different constants of the motion. 
Let us remark that as it was mentioned in Section II, the matrices $J(x,y)$ 
constructed according to (\ref{ans}) have rank $2$ hence, there are many Casimir 
functions which have vanishing Poisson bracket relations with any other 
dynamical quantity. A few words regarding the construction of Casimir functions 
for Poisson structures such as these seem in order. Time independent Casimir 
functions are, of course, constants of the motion. Consider now a different 
evolution along another parameter (call it $s$) which is given in terms of the 
symmetry vector ${\eta}^a$. In other words, consider
\begin{equation}
\frac{dx^a}{ds} = {\eta}^a(x^b)\ \,\ \ \ \ \ \ \ \ \ \ a,b = 1,........,N\ .
\label{eqmo1}
\end{equation}
Deformation of constants of the motion along ${\eta}^a$ may be viewed as
the evolution of such constants in the parameter $s$. A Casimir
function is such that its evolutions (both in time and in the $s$
parameter) vanish. So, the construction of Casimir functions may be
viewed as the search of entities which are simultaneously constant for
both the time and $s$ evolutions.

    To illustrate this, take for instance Eqs. (\ref{kdvk1}) and
(\ref{kdvk2}) and rewrite them as
\begin{equation}
\frac{dH_1}{ds}  = -  3 H_1 \ .
\label{kdvk1n}
\end{equation}
and
\begin{equation}
\frac{dH_2}{ds}= -  5 H_2\ .
\label{kdvk2n}
\end{equation}
solve them and eliminate the parameter $s$ to get that ${H_1}^{1/3}
{H_2}^{- 1/5}$ is a Casimir function for both of the Poisson matrices
(\ref{kdvj1}) and (\ref{kdvj2}). Similarly, one can get Casimir
functions for other Poisson matrices.

\section{Burgers Equation}

Consider the Burgers equation,
\begin{equation}
u_{t} = u_{xx} + 2uu_{x} \equiv f \ ,
\label{burgers}
\end{equation}
$x$ being in $[-a/2,a/2]$.
It is straightforward to prove that
\begin{eqnarray}
H_{1} &=& \int_{-a/2}^{a/2} dx  u(x)   \ ,
\label{burgersham1}  \\
H_{2} &=& \int_{-a/2}^{a/2} dx \exp\left[\int_{-a/2}^{x} u(y) dy
\right]  \ ,
\label{burgersham2}
\end{eqnarray}
are conserved quantities for it,
if we assume that the field vanishes at the boundary $\pm a/2$.

A symmetry transformation for Eq. (\ref{burgers}) is given by
\begin{equation}
\eta(x) = u(x)\exp\left[-\int_{-a/2}^{x} u \right]  \ .
\label{burgerssym}
\end{equation}
To see this, it is enough to show that
\[
{\cal L}_{f} \eta = \int_{-a/2}^{a/2} dy \left( \frac{\delta
f(x)}{\delta u(y)}\eta(y) - \frac{\delta \eta(x)}{\delta u(y)}
f(y) \right) = 0 \ .
\]
In fact,
\begin{eqnarray*}
 \int_{-a/2}^{a/2} dy \frac{\delta
f(x)}{\delta u(y)}\eta(y) &=&
\int_{-a/2}^{a/2} dy\frac{\delta \eta(x)}{\delta u(y)}
f(y) \\  &=& \exp\left( \int_{-a/2}^{x} u(y)dy \right) \left(
u_{xx} + uu_{x} - u^{3} \right) \ .
\end{eqnarray*}
The deformation of $H_{1}$ along $\eta$ does not vanish,
\begin{eqnarray*}
K_{1} = {\cal L}_{\eta} H_{1} &=& \int_{-a/2}^{a/2} dx\frac{\delta
H_{1}}{\delta u(x)}
\eta(x) \\
&=& \int_{-a/2}^{a/2} dx u(x)\exp \left(-\int_{-a/2}^{x} dw u(w)
\right) \\ &=& 1 - \exp (-H_{1}) \ .
\end{eqnarray*}
It may be proved that the  deformation of $H_{2}$ along $\eta$ is also non
vanishing,
\begin{eqnarray*}
K_{2} = {\cal L}_{\eta} H_{2} &=& \int_{-a/2}^{a/2} dx\frac{\delta
H_{2}}{\delta u(x)}
\eta(x) \\
&=& \int_{-a/2}^{a/2} dz \exp\left( \int^{z}_{-a/2}\right)
u(z) \left[\int_{z}^{a/2} dx \exp \left(\int_{-a/2}^{x} dw u(w)
\right) \right] \\ &=& H_2 - a \ .
\end{eqnarray*}
Therefore, we may construct Hamiltonian theories for the Burgers
equation using either $H_{1}$ or $H_{2}$ as Hamiltonians. The appropriate
Poisson matrices are
\begin{equation}
J_{1}(x,y) = \frac{f(x)\eta(y) - f(y)\eta(x)}{K_{1}} \ ,
\label{po1}
\end{equation}
and
\begin{equation}
J_{2}(x,y) = \frac{f(x)\eta(y) - f(y)\eta(x)}{K_{2}} \ ,
\label{po2}
\end{equation}
respectively.

 It is easy to check, for example, that,
\begin{equation}
C=\frac{e^{H_{1}}-1}{H_{2}-a} \ ,
\label{casim}
\end{equation}
is a Casimir for both of the Poisson brackets defined by (\ref{po1}) and
(\ref{po2}),
where we have used the approach described at the end of the preceding Section.

\section{Harry--Dym Equation}

The Harry-Dym equation is
\begin{equation}
u_{t} = \left(u^{-1/2}\right)_{xxx} \equiv f\ ,
\label{harry}
\end{equation}
where $x \in [-a/2,a/2] $.
If we assume periodic boundary conditions, $H_1$ and $H_2$
\begin{eqnarray}
H_{1} &=& \int_{-a/2}^{a/2} u dx  \  , \label{harryham1} \\
H_{2} &=& \int_{-a/2}^{a/2} u^{1/2} dx  \ , \label{harryham2}
\end{eqnarray}
are conserved quantities. Note that $a$ may be set equal to $\infty$.

Let us define the vector field
\begin{equation}
\xi(x) = Au - Bxu_{x}  \ ,
\end{equation}
where $A$ and $B$ are real constants. Let us now compute the Lie
derivative of it along $f$,
\begin{equation}
{\cal L}_{f} \xi  = 3 \left( \frac{1}{2}A + B \right)
\left(u^{-1/2}\right)_{xxx} = 3 \left( \frac{1}{2}A + B \right) f  \ .
\end{equation}
Therefore, $\eta$
\begin{equation}
\eta =-3 t \left( \frac{1}{2}A + B \right) f +\xi  \ ,
\end{equation}
is a symmetry transformation and can be used to construct a Hamiltonian
theory provided it deforms non trivially some Hamiltonian.

This is exactly the case for $H_{1}$ and $H_{2}$. In fact,
\begin{eqnarray}
K_{1} \equiv {\cal L}_{\eta} H_{1} &=&  (A+B) H_{1} \ , \\
K_{2} \equiv {\cal L}_{\eta} H_{2} &=& \left( \frac{A}{2} + B \right)
H_{2} \ ,
\end{eqnarray}
so we have one family of Poisson matrices associated with $H_{1}$ and
another
one associated with $H_{2}$. They are
\begin{equation}
J_{1}(x,y) = \frac{ \left(u^{-1/2}\right)_{xxx}\left(Au(y) - Byu_{y}
\right)
 -  \left(u^{-1/2}\right)_{yyy}\left(Au(x) - Bxu_{x} \right)}{K_{1}}
\ ,
\end{equation}
and
\begin{equation}
J_{2}(x,y) = \frac{ \left(u^{-1/2}\right)_{xxx}\left(Au(y) - Byu_{y}
\right)
 -  \left(u^{-1/2}\right)_{yyy}\left(Au(x) - Bxu_{x} \right)}{K_{2}}
\ ,
\end{equation}
respectively. Of course, we must be careful to choose $A$ and $B$ in
such a way that either $K_1$ or $K_2$ (or both) be non-vanishing.

\section{Conclusions}

We have constructed Hamiltonian structures, without using Lagrangians, for
several linear and non--linear systems of partial differential \mbox{equations}
(field theory) based on a recently devised method \cite{sh,sh1}, which needs
only of the knowledge of a constant of motion and a solution of the symmetry
equation. The examples include the heat equation which has, up to now, been
considered to be non--Hamiltonian \cite{salmon}. The structures found are
singular in the sense that there exist Casimir functions (which have vanishing
Poisson bracket with any dynamical variable). This feature is present in many
other Hamiltonian theories, and it is sometimes unavoidable (as it is the case
of gauge and constrained systems), and it also appears in Hamiltonian
descriptions of some fluids. We should stress that being able to produce a
Hamiltonian structure (albeit singular) for a system of differential equations
constitutes progress with respect to the situation of having no Hamiltonian
structure at all. More examples will be discussed in forthcoming articles.

\section{Acknowledgments}

The authors would like to thank M.P. Ryan, Jr. for many enlightening
discussions. This work has been supported in part by Centro de
Recursos Educativos Avanzados (CREA, Chile), Fondo Nacional de Ciencia
y Tecnolog\'{\i}a (Chile) grants 196--0866 and 396--0008, and a
bi--national grant funded by Comisi\'on Nacional de Investigaci\'on
Cient\'{\i}fica y Tecnol\'ogica--Fundaci\'on Andes (Chile) and Consejo
Nacional de Ciencia y Tecnolog\'{\i}a (M\'exico). A.G. is deeply
grateful to Fundaci\'on Andes for a Doctoral Fellowship. The
institutional support of a group of Chilean private companies
(EMPRESAS CMPC, CGE, CODELCO, COPEC, MINERA ESCONDIDA, NOVAGAS,
BUSINESS DESIGN ASSOCIATES, XEROX CHILE) is also recognized.

\end{document}